\documentclass[11pt,twoside]{book} 
\usepackage{asp2008s}
\usepackage{times}
\usepackage{lscape}
\usepackage{epsf}

\usepackage{graphicx}
\usepackage{amssymb}
\usepackage{longtable}
\usepackage[figuresright]{rotating}
\usepackage{multirow}

\newcommand{\simless}{\mathbin{\lower 3pt\hbox {$\rlap{\raise 5pt\hbox{$\char'074$}}\mathchar"7218$}}}

\renewcommand{\textfraction}{0.15} 
\renewcommand{\topfraction}{0.85} 
\renewcommand{\bottomfraction}{0.70}

\mainmatter

\newlength{\deftabcolsep}
\begin{document}

\renewcommand{\topfraction}{1.}
\renewcommand{\bottomfraction}{1.}
\renewcommand{\textfraction}{0.}
\renewcommand \thesection{\arabic{section}}
\renewcommand \thesubsection{\arabic{section}.\arabic{subsection}}
\renewcommand
\thesubsubsection{\arabic{section}.\arabic{subsection}.\arabic{subsubsection}}
\title{Star Formation in the Eagle Nebula}   %%% Fill in title
\author{Joana M. Oliveira}   %%% Fill in author names
\affil{School of Physical and Geographical Sciences, Lennard-Jones Laboratories,
Keele University, Staffordshire ST5 5BG, UK} %%% Fill in author
%affiliations

\begin{abstract} %%% Abstract to run on from here.
M16 (the Eagle Nebula) is a striking star forming region, with a complex
morphology of gas and dust sculpted by the massive stars in NGC\,6611. Detailed
studies of the famous ``elephant trunks'' dramatically increased our
understanding of the massive star feedback into the parent molecular cloud.
A rich young stellar population (2$-$3\,Myr) has been identified, from
massive O-stars down to substellar masses. Deep into the remnant molecular
material, embedded protostars, Herbig-Haro objects and maser sources bear
evidence of ongoing star formation in the nebula, possibly triggered by the
massive cluster members.
M\,16 is a excellent template for the study of star formation under the
hostile environment created by massive O-stars. This review aims at providing
an observational overview not only of the young stellar population but also of
the gas remnant of the star formation process.
\end{abstract}

%%% MAIN BODY OF TEXT GOES HERE. CONSULT "INSTRUCTIONS FOR AUTHORS USING
%%% LATEX2E MARKUP", SECTIONS 2.3-2.6 FOR HELP WITH EQUATIONS, FIGURES,
%%% AND TABLES.

%\section{}   %%% Top level section head (remove "%" symbol)
%\subsection{}   %%% Second level section head (remove "%" symbol)
%\subsubsection{}   %%% Lowest level section head (remove "%" symbol)
%\section*{}	%%% Unnumbered top level section head (remove "%" symbol)
%\subsection*{}   %%% Unnumbered second level section head (remove "%"
%symbol)

\section{Overview}

\begin{figure}[ht]
\begin{center}
\includegraphics[scale=0.65,angle=0]{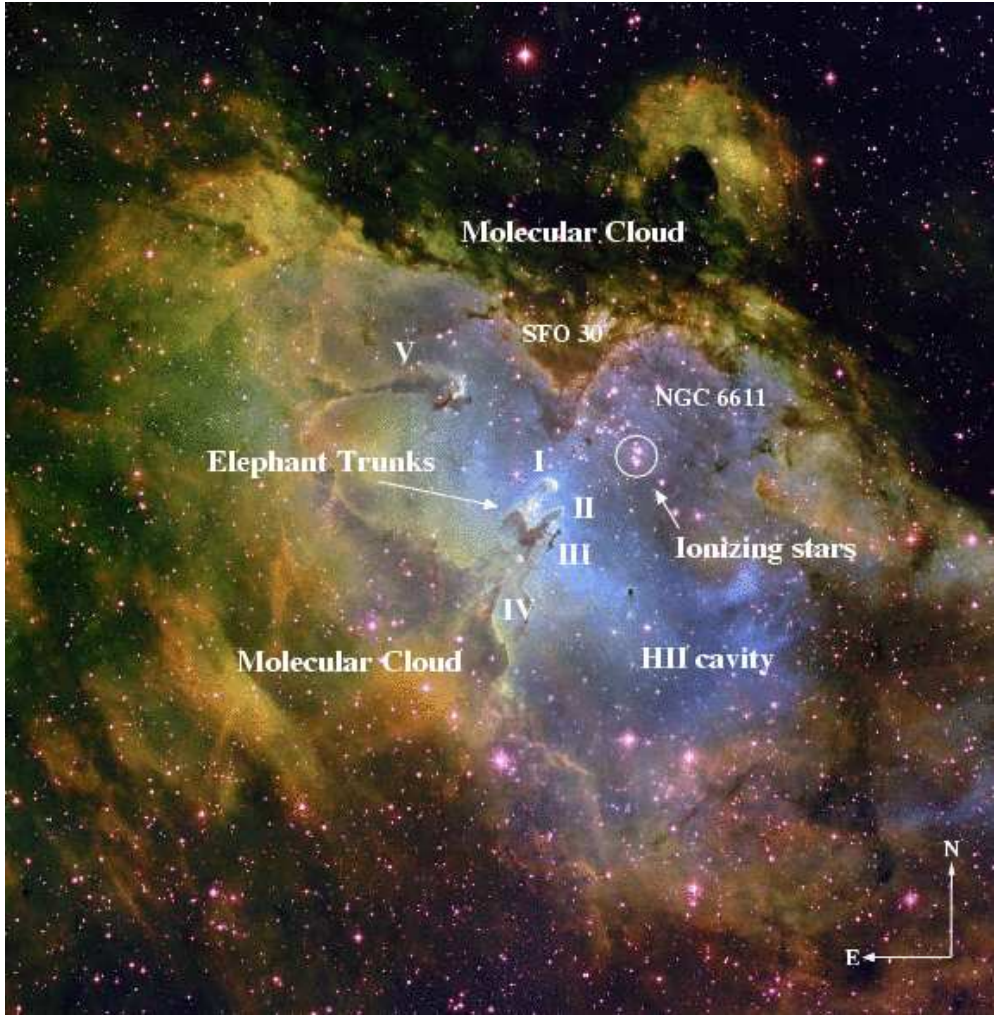}
\end{center}
\caption{Wide-field image (approximately 40\,arcmin$\times$41\,arcmin) of the
Eagle Nebula, showing the blister H\,{\sc ii} region created by the massive
stars in NGC\,6611. Towards the center of the image the dusty pillars are
clearly seen (labelled from I to V). This image was taken with the 0.9\,m
telescope at the Kitt Peak Observatory with the NOAO Mosaic CCD camera and it
combines H$\alpha$\, (green), [O\,{\sc iii}] (blue) and [S\,{\sc ii}] (red)
images. Credit: T.A.\,Rector, B.A.\,Wolpa and NOAO.}
\label{m16}
\end{figure}

The cluster Messier\,16 (M\,16), in the constellation Serpens Cauda, was first
discovered in 1745 by Jean-Philippe Loys de Cheseaux, a Swiss astronomer from
Lausanne, who in 1746 presented to the French Academy of Science a list of
clusters and nebulae, including M16. On June 3, 1764, Charles Messier
independently discovered the cluster, noting its nebulous nature, and gave it
the number M16 by which it is now known. It was later included in John
Herschel's catalogue of clusters and nebulae as h2006, and eventually received
the designation NGC\,6611 when listed in Dreyer's New General Catalogue. In the
popular literature, it is often known as the Eagle Nebula.

The massive stars in the young cluster NGC\,6611 \citep{walker61} are
responsible for the ionization of the H\,{\sc ii} region identified as Sh2$-$49
\citep{sharpless59}, Gum\,83 \citep{gum55} or RCW\,165 \citep*{rodgers60} in the
optical or W\,37 \citep{westerhout58} as observed at radio wavelengths. These
stars have been photoevaporating the surrounding parent molecular cloud and
sculpting the overdense molecular cores into the famous ``elephant trunks''
(Fig.\,\ref{m16}). First identified by \citet{duncan20}, these structures were
revealed in their full glory in the iconic HST images of \citet{hester96}; they
appear in the optical images as opaque fingers or columns of dense obscuring
material projected against the diffuse background nebular emission. Even if not
as dense as originally thought \citep*{thompson02}, these columns harbour small
protrusions on or near their surface, the EGGs, a fraction of which seem to be
associated with embedded Young Stellar Objects (YSOs), signs of a recent episode
of star formation \citep[e.g.][]{mccaughrean02}.

NGC\,6611 cluster members are distributed over a region of $\sim$14\,arcmin
radius, with a higher concentration in the largely unobscured 4\,arcmin
radius central area \citep{belikov00,kharchenko05}. The cluster contains a large
number of massive stars as well as a large population of pre-main sequence (PMS)
stars \citep{hillenbrand93}. These optically visible members of NGC\,6611, with
masses in the range 2 to 85\,M$_{\odot}$, have an average age of 2$-$3\,Myr
\citep{hillenbrand93,belikov00}. More recently a rich low-mass PMS population
(down to substellar masses) was identified by \citet{oliveira05} and
\citet{oliveira08}.

The Eagle Nebula is a superb example of the interaction of massive stars with
their environment, showcasing both its destructive power and its potential for
triggering star formation at the periphery of the remnant molecular cloud.

\section{The H\,{\sc ii} Region and the Molecular Cloud Structure}

Besides the study of the stellar population associated with NGC\,6611
(Section\,\ref{pop}), earlier work on M\,16 concentrated mainly on studying the
physical properties and kinematics of the H\,{\sc ii} region. Numerous H$\alpha$
and radio continuum maps of the region as well as spectra of hydrogen
recombination emission lines and OH and H\,{\sc i} absorption were obtained in
the 60's and 70's and the derived properties of the neutral and ionized gas are
summarized in \citet{goudis76} and references therein: radial and turbulent
velocities, densities and electron temperatures are provided in that paper. Many
of those early works also computed kinematic distances (Section\,\ref{dist}).

Radio spectra were also used to predict the properties of the ionizing stars in
NGC\,6611 \citep{goudis75}. \citet{felli70} resolve the ionized gas distribution
into 3 distinct continuum peaks, one in the northern obscured region
(Section\,\ref{sfo30}) and 2 either side of the main elephant trunks. The
ionized gas presents complex velocity fields, as shown by the splitting of the
optical emission line profiles into multiple velocity components
\citep*[][and references therein]{goudis76b,elliot78,mufson81}, related to the
interplay of the ionized flows with surrounding neutral material.

More recent observations of radio recombination lines (hydrogen, helium and
carbon) and radio continuum in M\,16 can be found in \citet{quireza06}.
\citet{garcia06} have measured the intensity of $\sim$\,250 emission lines in
bright regions in the nebula associated with NGC\,6611. This allowed them to
constrain the physical conditions in the H\,{\sc ii} region making use of a
variety of line intensity and continuum ratios. \citet{hebrard00} and
\citet{garcia06} also detected several deuterium Balmer lines and confirm that
fluorescence, not recombination, is the most probable hydrogen excitation
mechanism.

\subsection{The Elephant Trunks}

\begin{figure}[ht]
\begin{center}
\includegraphics[scale=0.275]{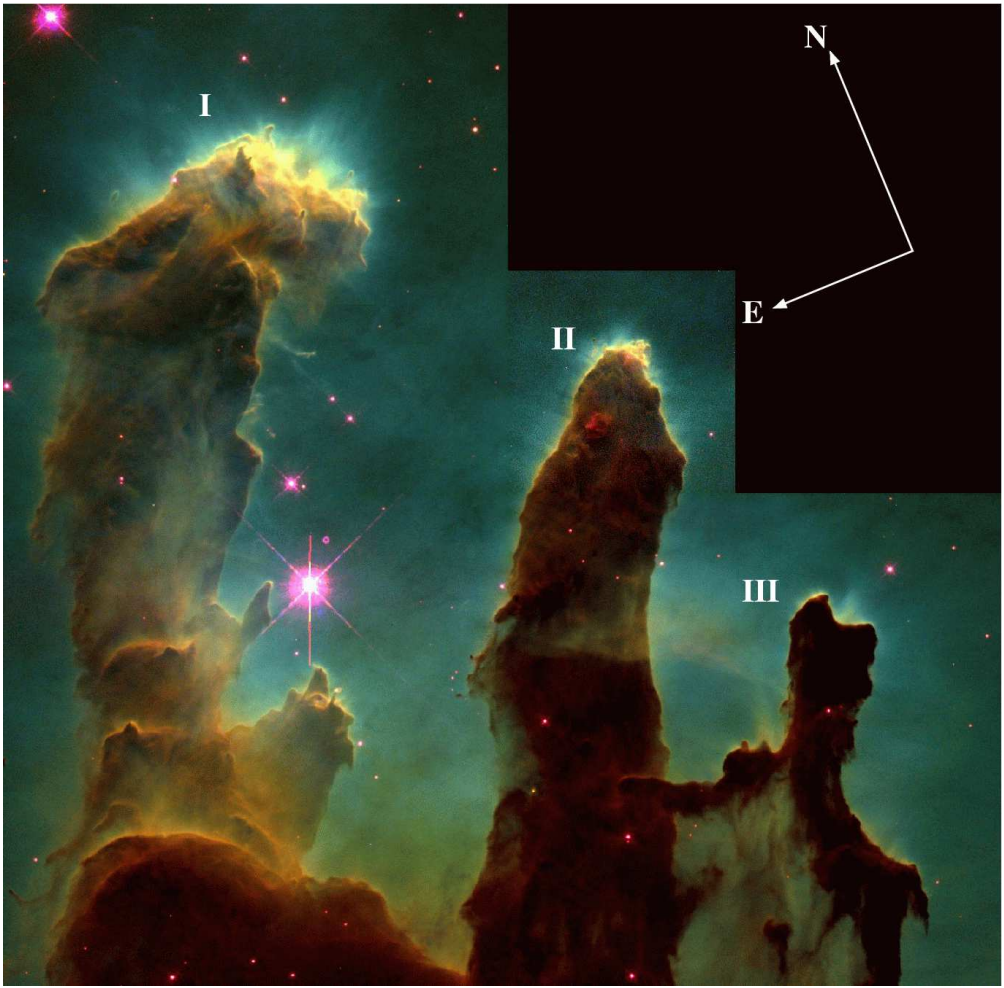}
\end{center}
\caption{HST WFPC2 colour composite image of a detail of the three central
elephant trunks in M\,16 \citep{hester96}. Columns are labelled I, II and III.
Red, green and blue colours show emission from [S\,{\sc ii}], H$\alpha$\, and
[O\,{\sc iii}] respectively. Credit: J.\,Hester, P.\,Scowen  and STScI.}
\label{trunks_o}
\end{figure}

\subsubsection{Morphology and Physical Conditions}
\begin{figure}[ht]
\begin{center}
\includegraphics[scale=0.67]{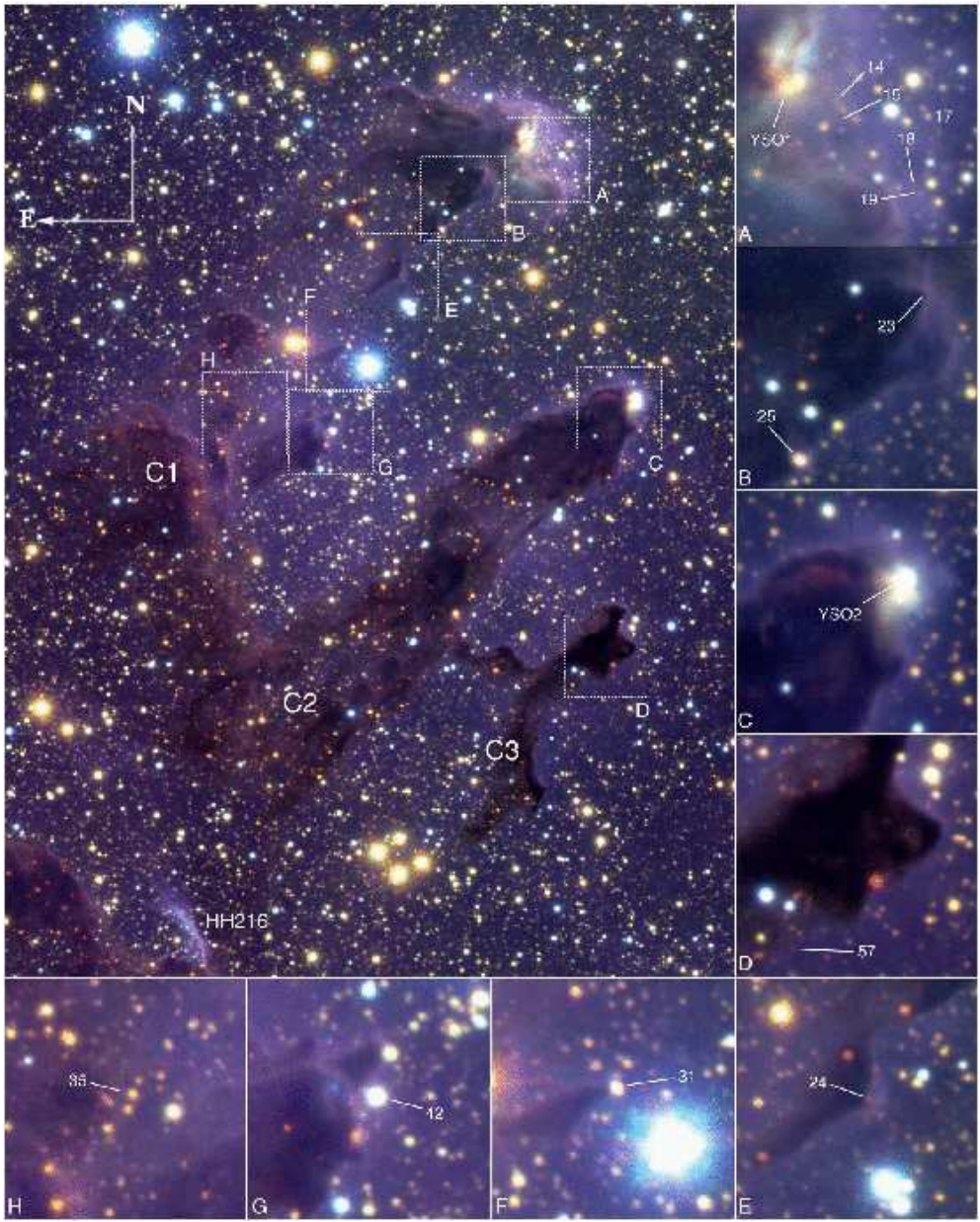}
\end{center}
\caption{True-colour near-IR image of the elephant trunks in M\,16
\citep{mccaughrean02}, obtained with ISAAC at the ESO/VLT. $JHK$ images are
shown respectively as blue, green and red. This image clearly demonstrates that
most of the molecular material in the columns sits at their tips. Insets show
detail of YSOs embedded in the molecular material, with the numbers referring to
the EGGs identified by \citet{hester96}. The location of the optically visible
Herbig-Haro object HH\,216 is also indicated (Section\,\ref{hh216}).}
\label{trunks_ir}
\end{figure}

The main elephant trunks in M\,16 are three dense structures of gas and dust
situated at the southeast ``boundary'' of the H\,{\sc{ii}} region created by
the numerous O-stars in NGC\,6611. These are normally referred to as columns I,
II and III respectively from northeast to southwest. \citet{hester96} used the
famous HST/WFPC2 images (Fig.\,\ref{trunks_o}) to investigate the morphology of
the columns and the ionization conditions at the interface between the
H\,{\sc{ii}} region and the molecular material. They found that both the
observed morphology and ionization structure are well understood in terms of
photoionization of the remnant molecular material. In very simple terms, the
ionizing radiation that reaches the molecular cloud increases the pressure at
the cloud interface, driving a photoevaporative flow of ionized material away
from the cloud into the (low density) H\,{\sc{ii}} region cavity
\citep[see][for a review]{hester05}. Lower density material is quickly blown
away, while denser molecular cores, compressed by the associated shock fronts,
survive longer. Near-IR images (\citealt*{sugitani02}; \citealt{thompson02})
reveal that, instead of being dense continuous structures of gas and dust,
columns I and II are made of relatively low-density material, capped by much
denser cores that effectively shield the columns from the ionizing radiation
(Fig.\,\ref{trunks_ir}). This shielding is thought to be responsible for the
finger-like morphology, with tails pointing away from the ionizing sources
\citep[e.g.][]{white99}. To the southeast of the elephant trunks, there
is another complex of molecular cloud material (column IV,
Fig.\,\ref{m16}), near the Herbig-Haro object HH\,216
\citep[e.g.][Section\,\ref{hh216}]{meaburn82a}. Several authors have used
hydrodynamics models to describe the formation and morphology of the Eagle
Nebula fingers \citep*[e.g.][]{williams01,miao06,mizuta06}.

Even at small scales, the structure sculpted by the advancing ionization front
is rather complex with numerous dense knots of gas and dust, known as
Evaporating Gaseous Globules (EGGs). These have been identified by
\citet{hester96} at, or near to, the surface of the three columns and have
typical sizes of the order of 300$-$400\,AU. It is thought that at least some
of these EGGs and the dense column caps are associated with ongoing star
formation (see next section).

Observations of the molecular gas in the pillars give further insight into their
physical properties. Maps of several CO isotopes probe the velocity field of the
molecular gas in the columns. \citet{pound98} and \citet{schuller06} found large
velocity gradients ($\sim$\,7.8\,km\,s$^{-1}$pc$^{-1}$) along the symmetry axis
of the pillars with position angles coincident with the direction of the nearby
O-stars, supporting the idea that such gradients are produced by the advancing
ionization front. \citet{white99} report on molecular line and continuum
emission observations. The dense submillimetre-continuum cores at the top of the
columns are constrained to have low dust temperatures ($\sim$\,10$-$20\,K),
cooler than the surrounding molecular gas ($\ga$\,60\,K). They estimate the
masses of these dense caps to be of the order of 10$-$60\,M$_{\odot}$,
approximately 55$-$80\% of the mass of each column, as measured from CO
observations; the total mass of the three columns is of the order of
$\sim$\,200\,M$_{\odot}$ \citep{white99}. Also using CO and millimetre continuum
observations, \citet*{fukuda02} propose a linear sequence of YSOs to starless
cores at the head of two of the columns, consistent with star formation activity
propagating along them.

Near-IR observations of both molecular and ionized hydrogen emission in the
elephant trunks \citep{allen99,levenson00} are useful probes of the physical
conditions in the Photo-Dissociation Region (PDR), the interface between the
molecular cloud and the H\,{\sc ii} region where the molecular hydrogen and dust
grains are dissociated by the far-UV radiation from the massive stars. Using
measured total Br$\gamma$ fluxes at the head of the columns and assuming that
the main ionizing stars are at a projected distance of 2\,pc, \citet{allen99}
find that their combined contribution to the local FUV field is
$\sim$\,16\,erg\,cm$^{-2}$\,s$^{-1}$, approximately 10$^{4}$ times the ambient
interstellar values.

Further evidence of the continuing destruction of the molecular cloud comes
from ISOCAM-CVF observations. Mid-IR spectra of the columns reveal emission due
to Unidentified Infrared Bands (commonly attributed to Polycyclic Aromatic
Hydrocarbons, PAHs) and atomic fine-structure line emission \citep{urquhart03},
both good tracers of the PDR conditions. \citet{urquhart03} find that UIB
emission is contained within the dusty pillars, specially around the tip of
column I. From the measured atomic fine-structure line ratios, they estimate
a surface ionizing flux of 1$-$3\,$\times$\,10$^{10}$
photons\,cm$^{-2}$\,s$^{-1}$
\citep[see also][]{white99}, consistent with the total ionizing flux computed by
\citet{allen99}.

\subsubsection{Young Stellar Objects}
By comparing optical HST and near-IR images, \citet{hester96} proposed that the
elephants trunks, in particular the EGGs, harbour YSOs and thus are active
sites of star formation. \citet{white99} suggested that cloud cores at the cap
of the pillars have temperature and density consistent with those expected in
the very early stages of protostellar formation. There is evidence for ongoing
star formation in the denser areas of M\,16 (maser emission, embedded YSOs) but
it is uncertain whether it is being triggered or uncovered by the ionization
front from the massive stars \citep{hester05}. \citet{indebetouw07} find that
the YSO distribution throughout the region supports a picture of distributed
low-level star formation, with no strong evidence for triggered star formation
in the pillars. Indeed, \citet{ogura02} also find no concentration of H$\alpha$
emission stars towards the trunks (section\,\ref{ha}).

Mid-IR observations (ISO/ISOCAM) revealed only a single embedded source,
associated with one of the EGGs, and \citet{pilbratt98} suggest that this hints
at a rather low level of on-going star formation in the columns. Instrumental
limitations, however, made this result inconclusive \citep{urquhart03}. Several
authors \citep{sugitani02,thompson02,mccaughrean02} describe embedded YSOs at
the tip of two of the columns --- these protostars are described in more detail
in Section\,\ref{individual}. Water maser emission
\citep[][Section\,\ref{masers}]{healy04} and Herbig-Haro objects
\citep[][Section\,\ref{hh216}]{andersen04} associated with the columns are
further evidence of ongoing star formation.

\citet{walsh82} identified 8 near-IR sources in the vicinity of the trunks, with
4 of those sources presenting colours consistent with emission from hot
circumstellar material. Based on a $JHK$ colour-colour diagram,
\citet{sugitani02} also identify a sample of Class II sources (young stars with
circumstellar dust disks) scattered throughout and around the three columns.
\citet{mccaughrean02} obtained high-resolution $JHK$ images that show the
elephant trunks in unprecedented detail (Fig.\,\ref{trunks_ir}). Of the 73 EGGs
they analysed, only $\sim$\,15\% have IR counterparts and are thus associated
with young low mass stars. As some EGGs remain opaque even at these wavelengths,
this IR-association rate must be considered as a lower limit. Assuming that the
observed J and H-band fluxes are photospheric, they derive masses and optical
extinction for these IR sources: 7 EGGs are associated with substellar masses
while 4 EGGs have masses in the range 0.35$-$1\,M$_{\odot}$. \citet{linsky07}
found that none of these 73 EGGs have X-ray counterparts (Section\,\ref{xrays}).
This lack of associated IR and X-ray emission seems to suggest that contrary
to what was proposed by \citet{hester96}, many of the EGGs do not contain
embedded YSOs.

\subsection{The Bright Rimmed Cloud SFO\,30}
\label{sfo30}

Bright Rimmed Clouds (BRCs) are dense clumps at the rim of molecular clouds
associated with (relatively) old H\,{\sc{ii}} regions. Following
\citet*{sugitani91}, BRCs are now commonly referred to as SFOs. SFO\,30 is
associated with the source IRAS\,18159$-$1346, situated in the optically
extincted region to the northeast of the ionizing stars in NGC\,6611 (also
called the north bay). \citet{morgan04} used archival radio, optical and IR
images to characterise the physical properties of many BRCs. They compare the
gas pressure of the ionized boundary layer and of the interior molecular gas and
conclude that SFO\,30 (and the majority of their sample) is in approximate
pressure equilibrium, and therefore it is likely that photoionization induced
shocks are propagating into the interior of the molecular cloud. Based on SCUBA
observations,  \citet{morgan08} found that many SFOs harbour star formation in
its early stages and SFO\,30 is the most luminous core in their sample
($\sim$\,7\,000\,L$_{\odot}$).

\citet{hillenbrand93} identified a population of embedded sources in near-IR
images towards this region. H$_2$O maser emission has also been detected
\citep*{healy04,valdettaro05} hinting at ongoing star formation in this area of
the molecular cloud. A tentative CS detection and failed ammonia detection have
been reported by \citet{anglada96}. One of the continuum peaks detected by
\citet{felli70} is also in this region.

\subsection{Column\,V}

\label{spire}
\begin{figure}[ht]
\begin{center}
\includegraphics[scale=0.52]{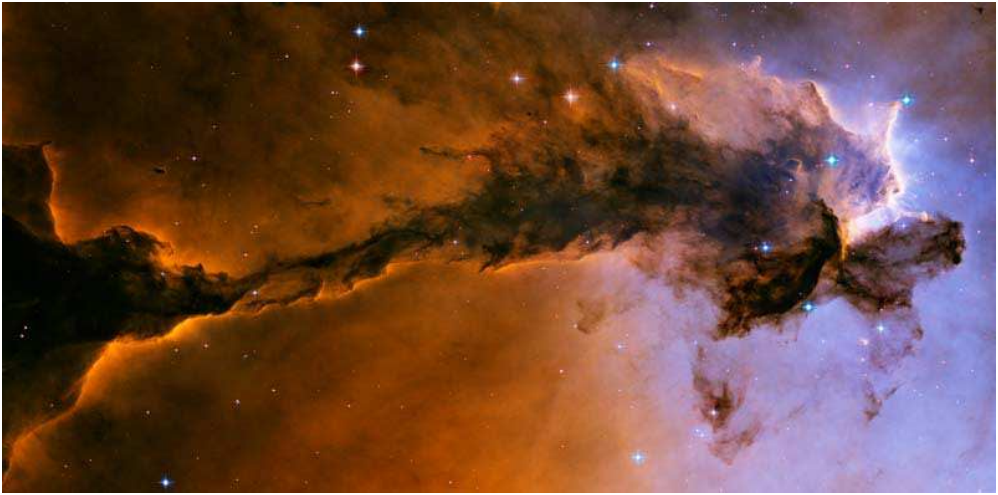}
\end{center}
\caption{Optical composite image of column\,V in M\,16, also known as ``the
spire''. These images were obtained with the instruments ACS and WFC on-board
HST. Star formation is occurring in this dusty pillar (Section\,\ref{ysos_5}).
Credit: NASA, ESA, and The Hubble Heritage Team (STScI/AURA).}
\label{spire_fig}
\end{figure}

To the northeast of the main dust pillars, there is another elongated column,
with a dense cap and a bright rim, identified as column\,V or ``the spire''
(Figs.\,\ref{trunks_o} and \ref{spire_fig}). \citet{meaburn86} identified a
high-speed ionized knot near its tip, that could be an Herbig-Haro object.
Multiple water maser emission components have been detected at the tip of this
column (\citealt{healy04}, Section\,\ref{masers}) as well as the MSX point
source G017.0335+00.7479 \citep{egan03} 5\,arcsec south of one of the maser
components. The ISOGAL survey has identified a bright YSO candidate
J181925.4$-$134535 at the base of the column \citep{felli02}. Recently,
\citet{indebetouw07}, using Spitzer Space Telescope IRAC and MIPS images
(Fig.\,\ref{spitzer}), identified the mid-IR counterparts of the maser emission
and constrained the properties of the IR source at the base of the column
(Section\,\ref{ysos_5}).

\subsection{The H\,{\sc ii} Region at Mid-IR Wavelengths}

\begin{figure}[p]
\begin{center}
\includegraphics[width=0.78\textwidth]{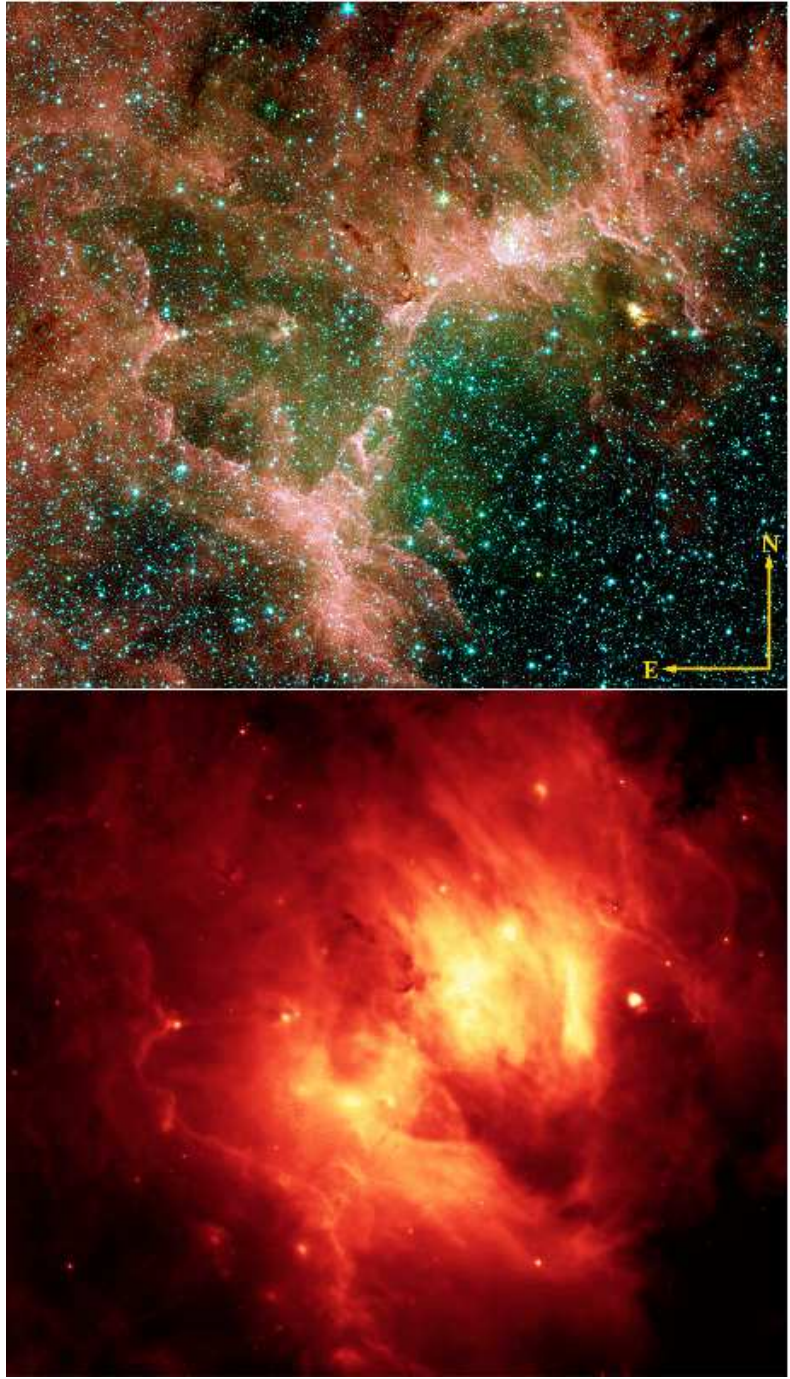}
\end{center}
\caption{Spitzer images of M\,16. Top: IRAC composite image, 3.6\,$\mu$m (blue),
4.5\,$\mu$m (green), 5.8\,$\mu$m (orange) and 8\,$\mu$m (red). Bottom: MIPS
24\,$\mu$m. Credit: NASA/JPL-Caltech/N. Flagey (IAS/SSC) \& A. Noriega-Crespo
(SSC/Caltech).}
\label{spitzer}
\end{figure}

Recently, Spitzer images (Fig.\,\ref{spitzer}) challenged once again our
understanding of M\,16. Fig.\,\ref{spitzer} (top) shows a composite of all IRAC
bands (3.6, 4.5, 5.8 and 8\,$\mu$m): the stellar population appears blue,
ionized gas green and hot dust and PAHs as red. Fig.\,\ref{spitzer} (bottom)
shows the 24~$\mu$m image. These images show that the cavity of the H\,{\sc ii}
region is full of warm dust. It has been suggested, based on the morphology of
the 24\,$\mu$m-emitting region, that a supernova explosion might be responsible
for heating the dust.

\section{Reddening and Distance Determinations}
\label{dist}

The determination of the distance to NGC\,6611 is complicated by the fact that
the extinction towards the stars in the cluster does not follow the standard
interstellar extinction law \citep[e.g.][]{sagar79}. Several authors have
studied the extinction properties towards NGC\,6611 cluster members
\citep{neckel81,chini83,the90,hillenbrand93,winter97,belikov99,yadav01,kumar04}
and they all agree that not only is the optical extinction ($A_{\rm V}$) very
patchy but the values of $R_{\rm V}$ (the ratio of total to selective
extinction), estimated on a star by star basis, seem to be larger than the
normal interstellar medium value, suggesting the presence of larger dust grains.
Indeed, polarimetric observations by \citet*{orsatti00,orsatti06} suggest the
presence of larger silicate and graphite grains, when compared with the standard
interstellar medium. Reported $R_{\rm V}$ values towards NGC\,6611 are in the
range 3.5$-$4.8 (typical value $\sim$\,3.75, \citealt{hillenbrand93}) while
$E(B-V)$ values vary between 0.5$-$1.1\,mag. Typical $E(B-V)$ towards the
cluster core is of the order of 0.75\,mag \citep{belikov99,dufton06}.

Distance determinations to NGC\,6611 are summarised in Table\,\ref{distance}.
\citet{walker61} first estimated the distance to NGC\,6611 to be
3.2\,$\pm$\,0.3\,kpc, a value also adopted by \citet{sagar79}. Early kinematic
distances were in the range 2.2$-$3.4\,kpc \citep{dieter67,mezger67,miller68}.
However, the kinematic distance obtained by \citet*{mcbreen82} places the
cluster considerably closer at 2.1\,$\pm$\,0.4\,kpc. \citet{belikov99} estimated
the distance by making use of all photometry then available, arriving at a value
of 2.14\,$\pm$\,0.1\,kpc. More recent determinations favour an even smaller
distance. \citet{dufton06} found a distance of 1.8\,$\pm$\,0.1\,kpc, based on
spectroscopic parallaxes for 24 massive cluster members. Recent determinations
using the main-sequence turnoff \citep*{bonatto06,guarcello07} are consistent
with this latest parallax determination.

\begin{table}[t]
\caption{Distance estimates to NGC\,6611.}
\label{distance}
\smallskip
\begin{center}
{\small
\begin{tabular}{|l|l|l|}
\hline
\multicolumn{1}{|c|}{distance} & \multicolumn{1}{c|}{reference} & \multicolumn{1}{c|}{comment}\\
\multicolumn{1}{|c|}{kpc}&&\\
\hline
3.2\,$\pm$\,0.3   & \citet{walker61,sagar79}& normal extinction law\\
1.7               & \citet{becker63}        & 3-colour photometry\\
3.4               & \citet{mezger67}        & kinematic distance\\
2.6               & \citet{dieter67}        & kinematic distance\\
2.2               & \citet{miller68}        & kinematic distance\\
2.1\,$\pm$\,0.4   & \citet{mcbreen82}       & kinematic distance\\
2.6\,$\pm$\,0.3   & \citet{the90}           & anomalous extinction law\\
2.0\,$\pm$\,0.1   & \citet{hillenbrand93}   & spectroscopic parallax\\
2.1\,$\pm$\,0.1   & \citet{belikov99}       & $UBV$ photometry\\
1.7               & \citet{kharchenko05}    & photometry\\
1.8\,$\pm$\,0.5   & \citet{bonatto06}       & $JH$ photometry\\
1.8\,$\pm$\,0.1   & \citet{dufton06}        & spectroscopic parallax\\
1.75              & \citet{guarcello07}     & $VI$ photometry\\
\hline
\end{tabular}}
\end{center}
\end{table}

\section{The Stellar Content of NGC\,6611}
\label{pop}

\subsection{High- and Intermediate-mass Population: Cluster Age and IMF}

\citet{walker61} reported that the intrinsic $V/(B-V)$ colour-magnitude diagram
of NGC\,6611 consists of a Main Sequence (MS), spectral types O5 to B5, with a
population of stars lying above it; he estimated an age of 1.8\,Myr for this
population. Several photometric studies of the high-mass population have been
carried out since, mainly to complete the census of massive stars in the region
and to constrain the anomalous reddening law and distance
\citep*[e.g.][]{hoag61,hiltner69,sagar79,the90,chini90,chini92}.
Proper motion studies have been performed by various authors
\citep*{schewick62,kamp74,tucholke86,hillenbrand93,kharchenko95,belikov99,baumgardt00,loktin03}.
The most recent determination by \citet{kharchenko05} lists
pm$_{\rm ra}$\,=\,1.60\,$\pm$\,0.33\,mas\,yr$^{-1}$ and
pm$_{\rm \delta}$\,=\,$-$0.35\,$\pm$\,0.48\,mas\,yr$^{-1}$.

\begin{figure}[!t]
\centering
\includegraphics[scale=0.87,bbllx=114bp,bblly=113bp,bburx=540bp,bbury=423bp,clip=]{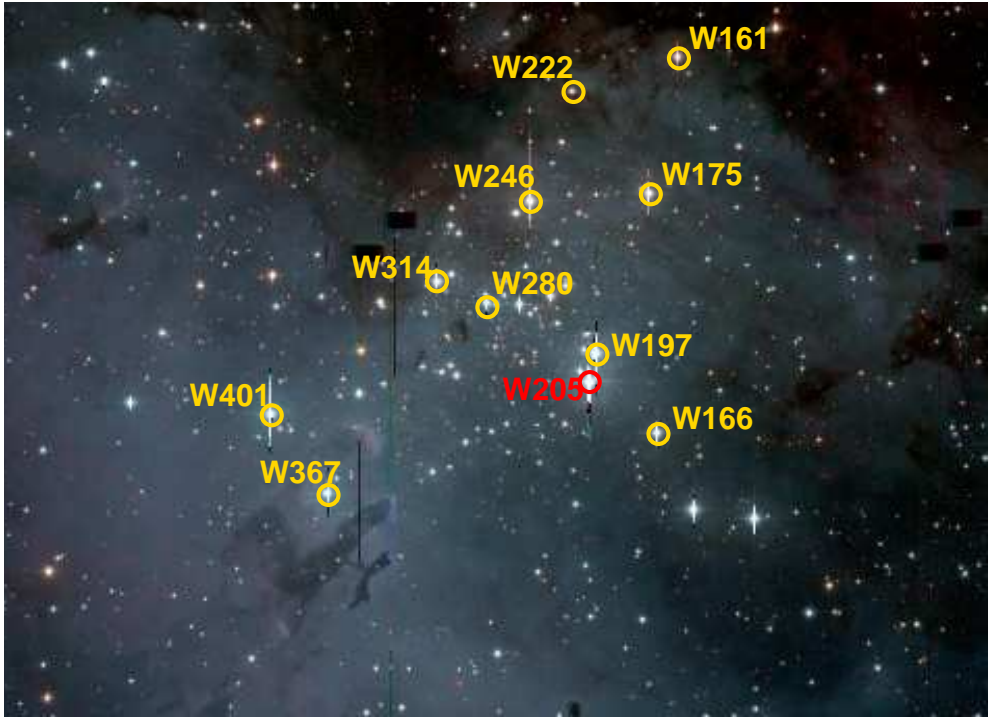}
\caption{ESO Imaging Survey (EIS) optical image of the Eagle Nebula
($\sim$\,15\,arcmin\,$\times$\,11\,arcmin across) with the O stars in
NGC\,6611 identified (see Table\,\ref{obstars}). W\,205 (O3$-$O5\,V) is shown in
red. Observations were carried out using the MPG/ESO 2.2\,m Telescope and the
ESO NTT at the La Silla Observatory.}
\label{ostars}
\end{figure}

The most massive star in NGC\,6611 is W\,205 or HD\,168076, with a mass of
75$-$80\,M$_{\odot}$ \citep{hillenbrand93,massey95,belikov00} and spectral type
O3$-$O5\,V \citep{hillenbrand93,evans05}. This star was identified early on as
the main ionizing source in the Eagle Nebula
\citep{gum55,sharpless59,walker61,gebel68,georgelin70,johnson73}, accounting for
about half the ionizing radiation in the Eagle Nebula \citep{hester96}. Massive
cluster members seem to have abundances consistent with early-type field
stars \citep*{brown86,daflon04} and no chemically peculiar stars had until
recently been found in the cluster \citep[][but see next section]{pauzen02}.
Several studies \citep[e.g.][]{bosch99,evans05} show that members' radial
velocities are in agreement with the radial velocity of the ionized gas in the
region \citep[$\sim$\,15\,km\,s$^{-1}$,][]{georgelin70}. \citet{duchene01}
conducted an adaptive optics survey of 96 stars in NGC\,6611, searching for
visual binaries. They found a high frequency of massive binaries
(18\,$\pm$\,6\%) and a lack of mass dependence in the binary properties. In
NGC\,6611, the binary properties seem to be more consistent with a canonical
accretion formation mechanism for OB stars \citep{beech94}, rather than the
merger mechanism \citep*{bonnell98}. A compilation of the O and B stars
(spectral types earlier than B3) in and towards M\,16 is provided in
Table\,\ref{obstars} and the positions of the O stars are shown in
Fig.\,\ref{ostars}.

\setlongtables
\LTcapwidth=8.2in
{\small
\begin{landscape}
\begin{longtable}[c]{|@{\hspace{1mm}}c@{\hspace{1mm}}|@{\hspace{1mm}}c@{\hspace{1mm}}c@{\hspace{1mm}}|@{\hspace{1mm}}r@{\hspace{1mm}}r@{\hspace{1mm}}r@{\hspace{2mm}}|@{\hspace{1mm}}c@{\hspace{1mm}}|@{\hspace{0.5mm}}r@{\hspace{1mm}}|c@{\hspace{1mm}}|l@{\hspace{1mm}}|}
\caption{Massive stars towards NGC\,6611. Identifications are from
\citet{walker61} and \citet{kamp74}. Positions (J2000) and $K$ magnitudes
are from 2MASS \citep{cutri03}; spectral types, $V$ magnitudes and $B-V$
colours are from \citet{evans05} and \citet{hillenbrand93}. Membership
probabilities (p), computed using positions and proper motions, are from
\citet{belikov99}. Additional references are 1: \citet{kumar04}; 2:
\citet{bosch99}; 3: \citet{orsatti00,orsatti06}; 4: \citet{duchene01}; 5:
\citet{evans05}; 6: \citet{alecian08}.\label{obstars}}\\
\hline
ID& RA ($^{h\,\,m\,\,s}$)&DEC ($^{d\,\,m\,\,s}$)   &\multicolumn{1}{c}{V}  &B-V& \multicolumn{1}{c|@{\hspace{1mm}}}{K}&SpT&p (\%)&other identifications&comments\\
%  &($^{h\,\,m\,\,s}$)&($^{d\,\,m\,\,s}$)& mag&mag&mag&   & \%        &&  \\
\hline
\endfirsthead
\caption{Massive stars in NGC\,6611 (continuation).}\\
\hline
ID&  RA($^{h\,\,m\,\,s}$)&DEC ($^{d\,\,m\,\,s}$)&\multicolumn{1}{c}{V}   &B-V&\multicolumn{1}{c|@{\hspace{1mm}}}{K}&SpT&p (\%)&other identifications&comments\\
%  &($^{h\,\,m\,\,s}$)&($^{d\,\,m\,\,s}$)& mag&mag&mag&   & \%        &&   \\
\hline
\endhead
\hline
\endfoot
\hline
\endlastfoot
W412&18:18:58.7&$-$13:59:28& 8.18&0.34& 7.39&B0 III	   &34&HD168183, BD$-$14$\deg$4991&spectroscopic binary$^{2}$\\
W205&18:18:36.4&$-$13:48:03& 8.18&0.43& 6.57&O4 V((f+))    &41&HD168076, BD$-$13$\deg$4926&visual binary$^{4}$\\
W197&18:18:36.1&$-$13:47:36& 8.73&0.45& 7.28&O6-7V((f))+B0 &46&HD168075, BD$-$13$\deg$4925&spectroscopic binary$^{2}$\\
&&&&& &&&&intrinsic polarisation$^{3}$\\
W401&18:18:56.2&$-$13:48:31& 8.90&0.04& 7.58&O8.5 V	   &46&HD168137, BD$-$13$\deg$4932&intrinsic polarisation$^{3}$\\
-   &18:20:34.1&$-$13:57:16& 9.13&0.44& 8.02&O8 III	   &$-$ &HD168504, BD$-$14$\deg$5005&\\
W367&18:18:52.7&$-$13:49:43& 9.39&0.24& 8.76&O9.7 IIIp     &76&BD$-$13$\deg$4930&\\
W468&18:19:05.6&$-$13:54:50& 9.40&0.28& 8.61&B0.5 V + B1:  &47&BD$-$13$\deg$4934&mid-IR excess$^{1}$; spectroscopic binary$^{5}$\\
 & & &  & &  & & & &intrinsic polarisation$^{3}$\\
W246&18:18:40.1&$-$13:45:18& 9.46&0.82& 6.70&O7 II(f)	   &88&BD$-$13$\deg$4927&\\
W503&18:19:11.1&$-$13:56:43& 9.75&0.49& 8.10&B1: e	   &40&BD$-$13$\deg$4936&HAeBe$^{1}$\\
W314&18:18:45.9&$-$13:46:31& 9.85&0.58& 7.94&O9 V	   &98&BD$-$13$\deg$4929&suspected spectroscopic binary$^{2,5}$\\
W150&18:18:30.0&$-$13:49:58& 9.85&0.48& 8.36&B0.5 V	   &86&BD$-$13$\deg$4921&intrinsic polarisation$^{3}$\\
W556&18:17:51.0&$-$13:50:56& 9.99&1.19& 6.26&B2.5I	   &40&BD$-$13$\deg$4912&\\
W125&18:18:26.2&$-$13:50:05&10.01&0.47& 8.60&B1 V + ?	   &79&BD$-$13$\deg$4920&\\
W175&18:18:32.7&$-$13:45:12&10.09&0.84& 7.00&O5 V((f+))+ late O   &95&BD$-$13$\deg$4923&visual binary$^{2}$; intrinsic polarisation$^{3}$\\
W280&18:18:42.8&$-$13:46:51&10.12&0.43& 8.76&O9.5 Vn	   &98&BD$-$13$\deg$4928&\\
W166&18:18:32.2&$-$13:48:48&10.37&0.57& 8.60&O9 V	   &95&&intrinsic polarisation$^{3}$\\
K601&18:19:20.0&$-$13:54:21&10.68&0.36& 9.40&B1.5 V	   &43&BD$-$13$\deg$4937&HAeBe star; strong magnetic field$^{6}$\\
W469&18:19:04.9&$-$13:48:20&10.69&0.40& 9.31&B0.5 Vn	   &70&BD$-$13$\deg$4933&HAeBe or classical Be$^{1}$;\\
&&&&& &&&&intrinsic polarisation$^{3}$\\
W254&18:18:40.8&$-$13:46:52&10.80&0.47& 9.28&B1 V	   &99&&visual binary$^{4}$\\
W483&18:19:06.5&$-$13:43:30&10.99&0.41& 9.61&B3 V	   &72&BD$-$13$\deg$4935&classical Be$^{1}$\\
W223&18:18:37.9&$-$13:46:35&11.20&0.59& 9.19&B1 V	   &89&&visual binary$^{4}$\\
W351&18:18:50.8&$-$13:48:13&11.26&0.45& 9.67&B1 V	   &$-$&&\\
W161&18:18:40.0&$-$13:43:08&11.29&1.05& 7.45&O8.5 V	   &70&&mid-IR excess$^{1}$; intrinsic polarisation$^{3}$\\
W536&18:19:18.5&$-$13:55:40&11.46&0.22&10.26&B1.5 V + ?    &39&&spectroscopic binary$^{5}$\\
W239&18:18:40.0&$-$13:54:33&11.48&0.36&10.16&B1.5 V	   &67&&\\
W210&18:18:37.0&$-$13:47:53&11.50&0.54& 9.99&B1 V	   &92&&\\
W259&18:18:41.0&$-$13:45:30&11.56&0.73& 9.19&B0.5 V	   &87&&intrinsic polarisation$^{3}$\\
W343&18:18:49.3&$-$13:46:51&11.72&0.85& 8.90&B1V	   &83&&\\
W296&18:18:44.7&$-$13:47:56&11.81&0.49&10.10&B1.5 V	   &95&&\\
W584&18:18:23.6&$-$13:36:28&12.02&1.05& 8.58&O9 V	   &47&&\\
W207&18:18:36.7&$-$13:47:33&12.07&0.53&10.17&B1V	   &89&&\\
W290&18:18:44.9&$-$13:56:22&12.12&0.46&10.89&B2.5 V	   &52&&\\
W275&18:18:42.2&$-$13:46:52&12.12&0.46&10.53&B1.5V	   &0&&visual binary$^{4}$; intrinsic polarisation$^{3}$\\
W301&18:18:45.0&$-$13:46:25&12.19&0.58&10.30&B2 V	   &92&&\\
W289&18:18:44.1&$-$13:48:56&12.60&0.52&10.90&B3 V	   &95&&\\
W444&18:19:00.4&$-$13:42:41&12.65&0.92& 9.74&B1.5 V	   &70&&\\
W300&18:18:45.0&$-$13:47:47&12.69&0.52&10.58&B1.5V	   &96&&\\
W231&18:18:38.5&$-$13:45:56&12.71&0.75&10.06&B1V	   &78&&\\
W311&18:18:45.6&$-$13:47:53&12.78&0.55&10.77&B3 V	   &$-$&&visual binary$^{4}$; intrinsic polarisation$^{3}$\\
W227&18:18:38.4&$-$13:47:09&12.83&0.62&10.60&B2 V	   &89&&visual binary$^{4}$\\
W472&18:19:04.7&$-$13:44:44&12.85&0.53&11.40&B3 V + ?	   &12&&spectroscopic binary$^{5}$\\
W409&18:18:57.4&$-$13:52:12&12.89&0.41& 9.91&B2.5 V	   &66&&intrinsic polarisation$^{3}$\\
W297&18:18:44.5&$-$13:45:48&12.89&0.67&10.41&B2 Vn	   &10&&visual binary$^{4}$; intrinsic polarisation$^{3}$\\
W25 &18:18:09.3&$-$13:46:54&12.93&0.98& 9.49&B0.5V	   &50&&visual binary$^{4}$; intrinsic polarisation$^{3}$\\
W267&18:18:41.7&$-$13:46:44&13.11&0.52&11.37&B3 V	   &87&&visual binary$^{4}$\\
W222&18:18:37.5&$-$13:43:39&13.08&1.35& 8.20&O7 V((f))     &88&&\\
W188&18:18:33.7&$-$13:40:58&13.13&1.34& 8.78&B0V	   &68&& visual binary$^{4}$\\
W541&18:19:19.1&$-$13:43:52&13.21&0.64&11.19&B1-3 V	   &56&&\\
W251&18:18:40.4&$-$13:46:16&13.34&0.69& $-$ &B2V	   &0&&\\
W371&18:18:53.0&$-$13:46:45&13.44&0.65&11.16&B0.5V	   &72&&\\
W305&18:18:45.0&$-$13:45:25&13.51&1.07& 9.80&B1 	   &84&&\\
W228&18:18:38.1&$-$13:44:25&13.51&0.93&10.08&B2V	   &86&&\\
W80 &18:18:18.2&$-$13:41:59&13.82&1.64& 9.30&B2V	   &24&&\\
W269&18:18:41.6&$-$13:42:47&13.98&0.93&10.65&B1.5V	   &26&&\\
\hline
\end{longtable}
\end{landscape}
}

\begin{figure}[ht]
\begin{center}
\includegraphics[scale=0.5]{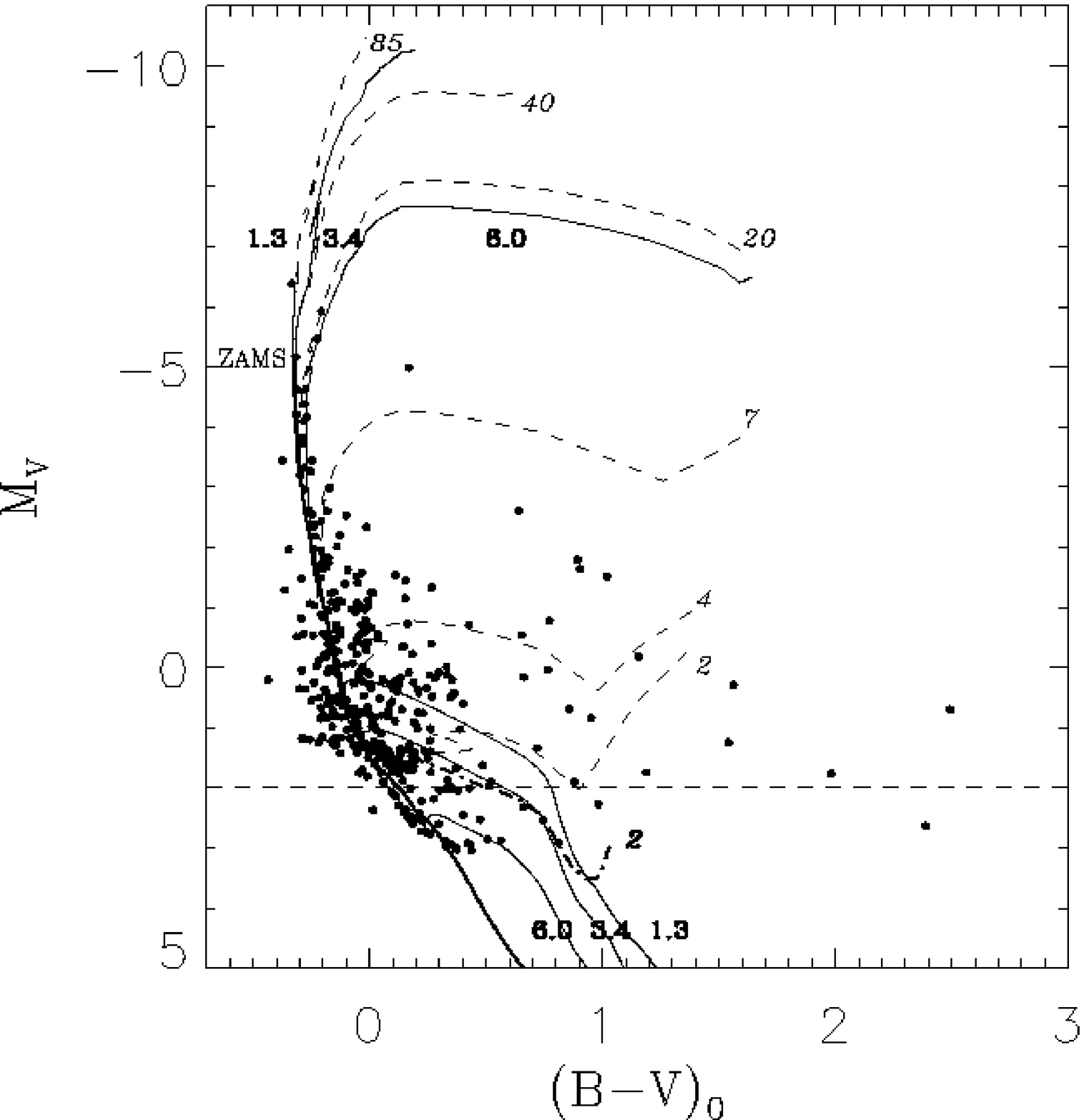}
\end{center}
\caption{Colour-magnitude diagram of the NGC\,6611 cluster members reproduced
from \citet{belikov00}. The ZAMS \citep{schmidt82} is shown by a thick line.
Thin lines (with bold age labels) are PMS isochrones \citep[1.3, 3.4 and
6\,Myr,][]{palla93}, while the dotted-dashed line at the bottom of the plot is
the corresponding PMS track for 2\,M$_{\odot}$. Post-MS evolutionary tracks
\citep{schaller92} are also plotted (dashed lines, with italic mass labels).
The PMS population can clearly be seen in this diagram.}
\label{cmd_belikov}
\end{figure}

Only in the last decade real progress was possible in the study of the cluster
PMS. There is a significant population of intermediate-mass stars
(2$-$8\,M$_{\odot}$) that sit well above the Zero-Age Main-Sequence (ZAMS) in
the Hertzsprung-Russell diagram
\citep[][ see also the colour-magnitude diagram in Fig.\,\ref{cmd_belikov}]{hillenbrand93,belikov00}.
The slope of the Initial Mass Function (IMF;
$\Gamma = {\rm d} \log \xi(\log m)/ {\rm d}  \log m$, \citealt{scalo86}) in
NGC\,6611 is broadly consistent (down to 2\,M$_{\odot}$) with what is found in
other Galactic high-mass star forming regions \citep{massey95}, with estimates
of $\Gamma$ in the range $-$1.5 to $-$0.7
\citep*{pandey92,hillenbrand93,massey95,belikov00,dufton06}. Recently,
\citet{bonatto06} analysed the population of NGC\,6611 down to 5\,M$_{\odot}$
and found that the mass function in the cluster core
($R_{\rm core} \sim 0.7 \pm 0.1$\,pc\footnotemark
\footnotetext{\citet{guarcello07} estimate the cluster core radius to be twice
this value, $1.4 \pm 0.08$\,pc.}) is relatively flat with a slope
$\Gamma = -0.62$, while in the halo (to an outer radius of
$R_{\rm lim} \sim 6.5 \pm 0.5$\,pc) it steepens to a slope of $\Gamma = -1.52$;
the overall IMF is similar to a Salpeter IMF with a slope of $\Gamma = -1.45$.
They propose that this spatial variation of the IMF slope may be a consequence
of mass segregation in the cluster. This steeper IMF is also consistent with a
slope of $\Gamma = -1.5$ as determined by \citet{dufton06}. The lower-mass IMF
is described in Section\,\ref{low_imf}.

Considering only the known cluster members more massive than 5\,M$_{\odot}$, the
lower-limit to the observed mass of the cluster is
(1.6\,$\pm$\,0.3)\,$\times$\,10$^{3}$\,M$_{\odot}$ \citep{bonatto06}. Assuming
that the stars in the mass range 6\,$-$\,12\,M$_{\odot}$ constitute 5.5\% of the
total mass of the complete population of stars spanning the range
0.1\,$-$\,100\,M$_{\odot}$, \citet{wolff07} estimate the total cluster mass to
be $\sim$\,25\,$\times$\,10$^{3}$\,M$_{\odot}$ with a density
28.5\,M$_{\odot}$\,pc$^{-3}$ \citep[see also][]{weidner06}.

The typical age of the NGC\,6611 population is of the order of 2$-$3\,Myr but a
considerable age spread ($<$\,1$-$6\,Myr) seems to be present
\citep{hillenbrand93,winter97,belikov00,dufton06}. \citet{bonatto06} favour a
younger age of 1.3\,$\pm$\,0.3\,Myr.

\citet{belikov99} compiled most of the then available astrometry and photometry
on NGC\,6611. From their catalogue of over 2000 objects (available on-line via
CDS), \citet{belikov00} use astrometric, photometric and proper motion criteria
to select 376 probable cluster members (Fig.\,\ref{cmd_belikov}) with masses
down to 2\,M$_{\odot}$.

\subsection{H$\alpha$\, Emission Stars}
\label{ha}

Optical photometry and proper motion methods have been commonly used to identify
members of NGC\,6611 (see previous section). Another classical way to search for
young stars in star forming regions are H$\alpha$\ emission surveys. Both Herbig
AeBe (HAeBe) and Classical T\,Tauri stars (CTTS) show Balmer line emission, in
particular in H$\alpha$. However, not all young intermediate- and low-mass young
stars show emission at all times. It is thought that strong and broad H$\alpha$
emission in young stellar objects originates from the interaction of the star
with its circumstellar disk, while weak H$\alpha$ emission originates solely
from activity in the stellar chromosphere in young stellar objects without a
disk \citep[e.g.][]{white03}. Furthermore H$\alpha$ emission is known to be
intrinsically variable \citep[e.g.][]{guenther97}. Taking these considerations
into account, H$\alpha$ emission surveys do not detect all low- and
intermediate-mass stars in a star forming region; however, particularly in
such young clusters like NGC\,6611, they are a useful tool to probe the young
stellar population.

In bright H\,{\sc ii} regions like the Eagle Nebula, contamination by the very
strong background H$\alpha$\ emission is a problem for slit and fibre
spectroscopy. As a result, a few HAeBe candidates were identified by
\citet{hillenbrand93} and \citet{winter97} (2 and 4 objects respectively), but
only four of these were confirmed with slitless grism spectroscopy by
\citet{herbig01}. \citet*{ogura02} performed a similar survey covering a much
larger area, including SFO\,30 and much of the elephant trunks. They identified
82 H$\alpha$\ emission objects, likely HAeBe and CTTS candidates, distributed
throughout the observed area, with no hint of concentration towards the remnant
molecular cloud --- positions and measurements for these objects are available
online via CDS. \citet{kumar04} have identified 4 HAeBe candidates based on
spectral properties and/or mid-IR excesses. For the HAeBe fast rotator  K\,601
(or W\,601), \citet{alecian08} report the detection of a strong magnetic field
and the presence of chemical peculiarities in its atmosphere.

\subsection{Low-mass Stars and Brown Dwarfs: IMF and Disk Population}
\label{low_imf}

\begin{figure}[t]
\begin{center}
\includegraphics[scale=0.65]{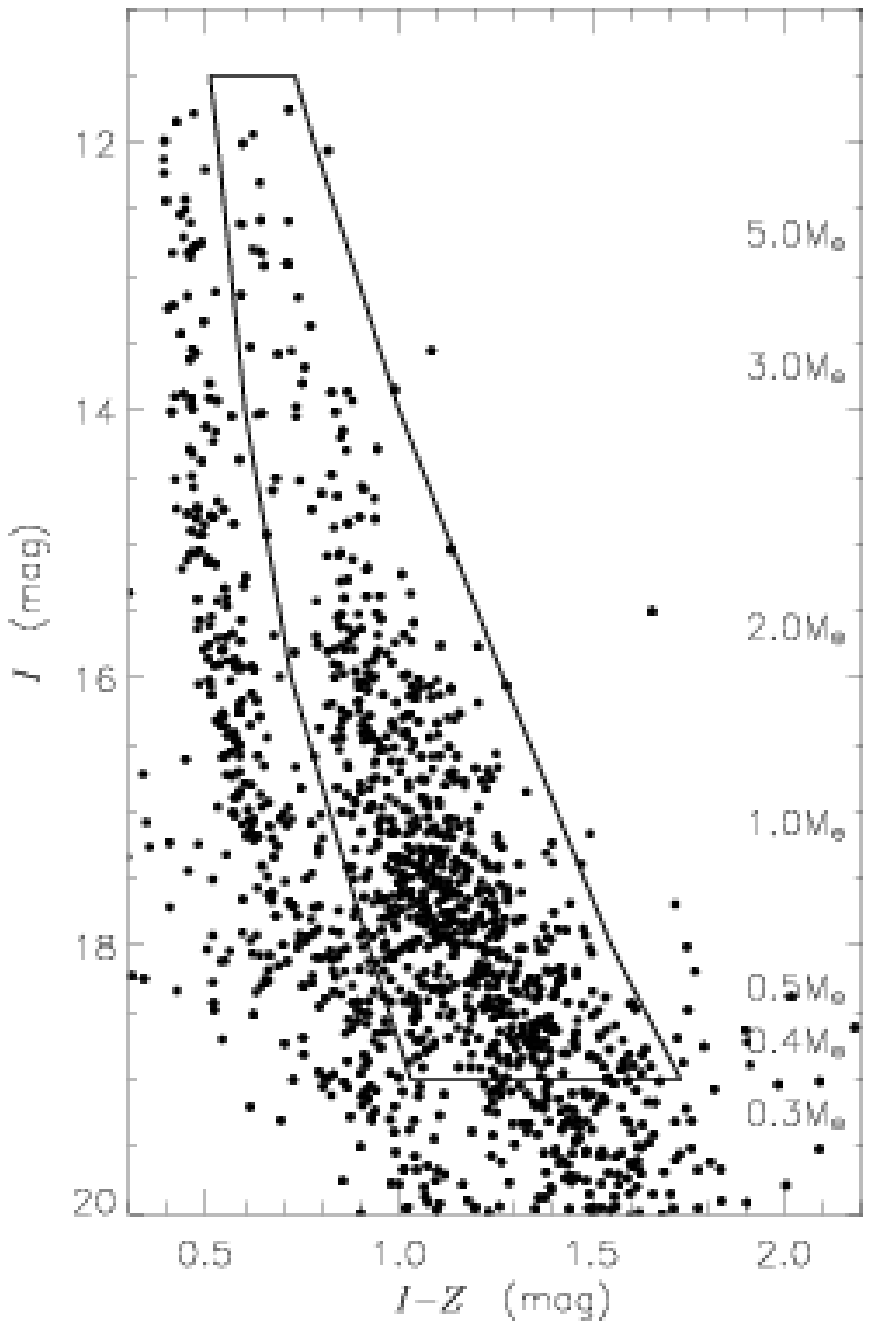}
\includegraphics[scale=0.65]{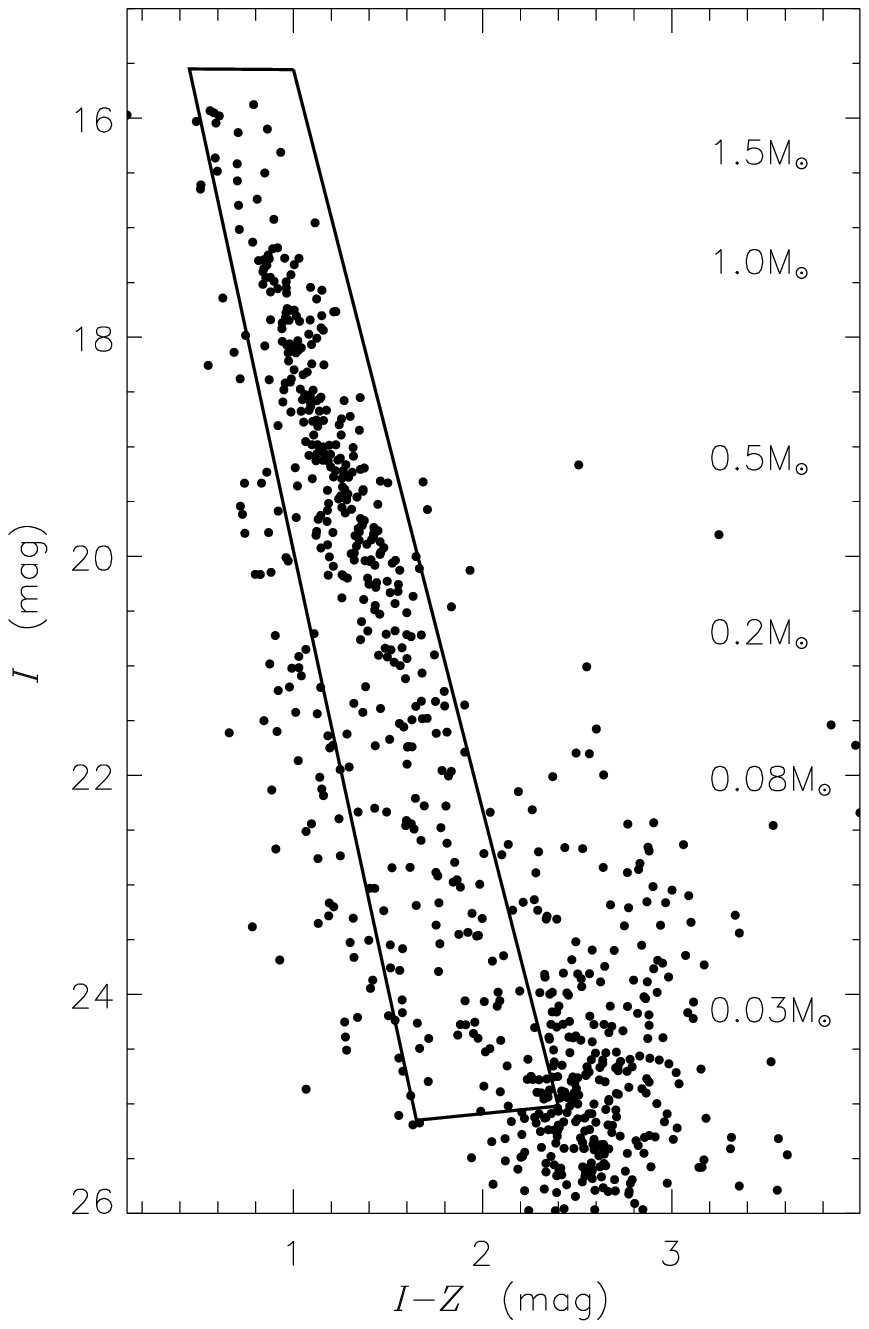}
\end{center}
\caption{Left: Colour-magnitude diagram for the central
7\,arcmin\,$\times$\,6\,arcmin region of NGC\,6611 \citep{oliveira05}. The PMS
can be clearly seen, separated from the bulk of field star contamination.
Approximate PMS masses are indicated \citep*{siess00} for a distance 1.8\,kpc,
an age of 3\,Myr and $E(B-V) =0.7$\,mag (Sect.\,\ref{dist}). Right: HST
colour-magnitude diagram of the central 2.5\,arcmin$^2$ area
\citep*{oliveira08}. Approximate PMS masses are indicated \citep*{baraffe98}.
The clump of red stars ($2 \la I-Z\la 3$) are evolved stars towards the Galactic
Center. Note that the ground-based and HST filter sets are very different and no
attempt was made to harmonize the magnitudes.}
\label{iz_cmd}
\end{figure}

Only recently has it become possible to probe the low-mass and brown dwarf
populations of NGC\,6611. \citet{oliveira05} performed an $IZ$ survey of the
cluster, with complementary $JHKL'$ imaging covering the central
4\,arcmin\,$\times$ 5\,arcmin area. In the optical colour-magnitude diagram the
rich low-mass PMS can easily be identified down to masses of
$\sim$\,0.5\,M$_{\odot}$ (Fig.\,\ref{iz_cmd} left). \citet*{oliveira08} used HST
to extend the study of the PMS population to well bellow the substellar
boundary, to about 0.02\,M$_{\odot}$ (Fig.\,\ref{iz_cmd} right). These
observations add several hundreds of cluster candidates to the published
samples, reaching for the first time into the brown dwarf regime.

\begin{figure}[t]
\begin{center}
\includegraphics[scale=1]{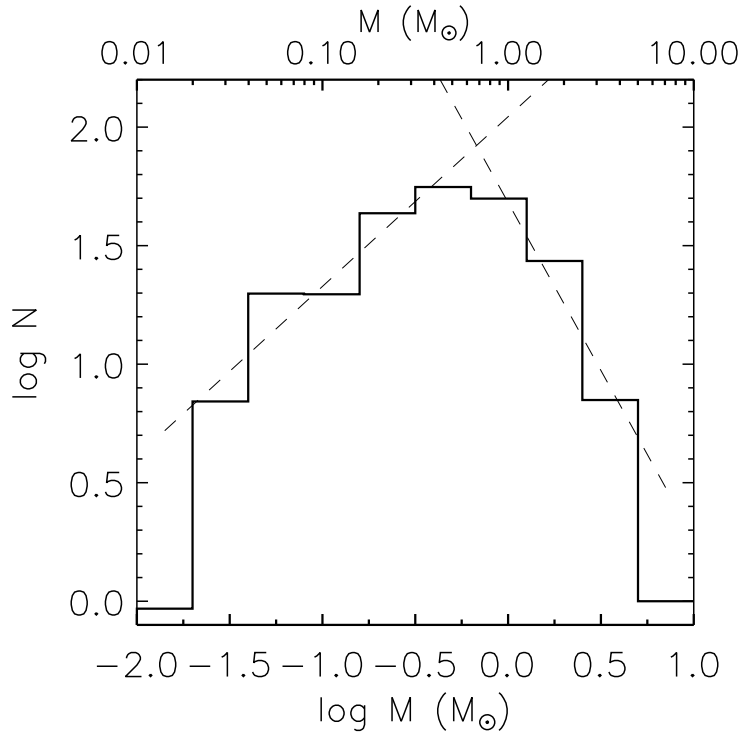}
\end{center}
\caption{Initial Mass Function for the central region of NGC\,6611
\citep*{oliveira08}. Power-law IMF slopes of $\Gamma = -$1.4 and $\Gamma = 0.7$
are plotted for reference \citep{kroupa01}.}
\label{imf}
\end{figure}

\citet*{oliveira08} constructed the IMF for the central 2.5\,arcmin$^2$ region
of NGC\,6611 (Fig.\,\ref{imf}) using the datasets described above. In agreement
with published work described previously, for higher masses the slope of the IMF
is consistent with a Salpeter index. For lower masses the IMF seems to flatten
between $\sim 0.3 - 1$\,M$_{\odot}$, with a peak at $\sim 0.4-0.5$\,M$_{\odot}$,
and then drop into the brown dwarf regime (slope $\Gamma \sim 0.7$), consistent
with a power-law IMF as described by \citet{kroupa01}. The brown dwarf to star
ratio of $\sim$\,0.13 is consistent with other star forming regions
\citep{luhman03}. For a full discussion of the IMF in NGC\,6611 refer to
\citet*{oliveira08}.

\begin{figure}[t]
\begin{center}
\includegraphics[scale=0.65]{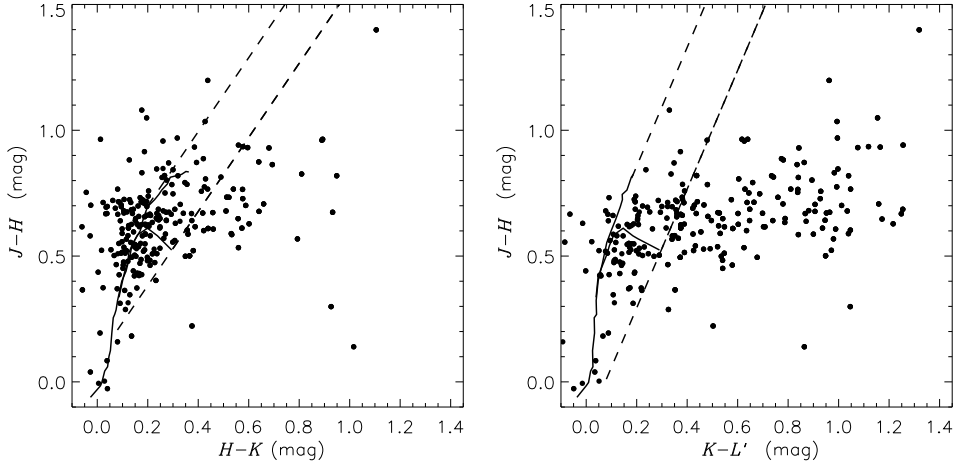}
\end{center}
\caption{Intrinsic colour-colour diagrams for PMS objects in NGC\,6611,
$JHK$ (left) and $JHKL'$ (right), from \citet{oliveira05}.
The loci of main-sequence and giant stars are represented (solid lines), as well
as the reddening band (dashed lines). Objects to the right of the reddening band
have an $H-K$ or $K-L'$ excess indicative of the presence of a dusty
circumstellar disk. }
\label{jhkl_ccd}
\end{figure}

Near-IR photometry was used by \citet{oliveira05} to investigate the population
of young low-mass stars that retain their circumstellar disks (i.e.\ CTTS).
Circumstellar dust disks are commonly found around very young low-mass stars.
If a young PMS star exhibits an IR colour in excess of the stellar photosphere
this indicates the presence of a circumstellar disk. The cluster colour-colour
diagrams are shown in Fig.\,\ref{jhkl_ccd}: the $JHK$ diagram diagnoses mainly
massive accretion disks \citep[e.g.][]{hillenbrand98} while the $JHKL$ diagram
detects the large majority of circumstellar dusty disks at this young age
\citep[e.g][]{haisch01}. \citet{oliveira05} found that the disk frequency in the
inner area of NGC\,6611 (50$-$60\%) is consistent with similarly aged but
quieter (i.e.\ with no O-stars) nearby star forming regions
\citep*{haisch01b}.

\section{X-ray Emission in M\,16}
\label{xrays}

\citet{guarcello07} and \citet{linsky07} analyse the same Chandra X-ray
Observatory images of NGC\,6611; the ACIS-I detector with
16.9\,arcsec\,$\times$\,16.9\,arcsec field-of-view covers a large fraction of
M\,16, including the HST-identified columns and the denser cluster region of
NGC\,6611. \citet{guarcello07} use the X-ray catalogue combined with 2MASS
K-band photometry to claim that there is a deficit of young stars with
circumstellar disks (as indicated by a K-band excess) towards the cluster
centre, possible evidence of the destructive power of the massive stars in the
cluster. However, as discussed in the previous section, this method does not
identify the majority of objects with circumstellar disks.

\citet{linsky07} have concentrated on the region associated with the dusty
pillars. They find that the vast majority of X-ray sources in this region are
moderately reddened young stars with small near-IR excesses. Only two hard X-ray
sources are identified with embedded protostars near or within the dusty pillars
--- these sources are discussed in more detail in Section\,\ref{individual}.
None of the EGGs analysed by \citet{mccaughrean02} have associated X-ray
emission. Of the 11 EGGs with IR counterparts, 7 have substellar masses and thus
are expected to be below the survey's detection threshold. More unexpectedly,
the 4 EGGs with masses 0.35$-$1.0\,M$_{\odot}$ are also not detected in the
X-ray survey and \citet{linsky07} conclude that these EGGs do not emit X-rays at
the level that one would expect for T\,Tauri stars. It is possible that these
objects are at the very early stages of YSO evolution and have not yet become
X-ray active. For the majority of the EGGs, however, both the lack of IR and
X-ray associated sources seems to hint that many of the EGGs do not contain
embedded YSOs.

\section{Water Maser Emission in M\,16}
\label{masers}

Water maser emission sources found in the molecular gas that surrounds
H\,{\sc ii} regions are a clear indication of objects in the earliest stages of
formation. Observational studies suggest that maser activity is associated with
YSOs of all luminosities and it normally lies in close proximity to the
exciting object \citep[e.g.][]{tofani95,furuya01}. Thus, water masers are
powerful probes of embedded star formation, even if their intrinsic variability
\citep{palagi93,tofani95,felli07} means that not all embedded YSOs are uncovered
by this type of observations.

Water maser emission in the Eagle Nebula was first identified by
\citet{yngvesson75} and \citet{blitz79} towards W\,37. The nature of maser
observations implies that surveys of complete star forming regions are normally
impractical. Thus, one normally searches for maser emission in locations where
star formation is suspected, for instance by the presence of embedded IR
sources. As a result, several maser sources associated with IRAS sources were
identified in M\,16: IRAS\,18156$-$1343 \citep{codella94}, IRAS\,18152$-$1346
\citep{braz83,codella95} and IRAS\,18159$-$1346
\citep[associated with SFO\,30,][]{braz83,healy04,valdettaro05} ---
\citet{molinari96} searched for ammonia towards IRAS\,18156$-$1343 but no
emission was detected. No maser emission was detected towards two other
IRAS sources in the region, IRAS\,18164$-$1340 and IRAS\,18160$-$1339
\citep*{szymczak00}. All these sources are located in the dense region to the
north of the massive stars in the cluster. Recent surveys have not detected
maser emission towards W\,37 as initially reported; this is probably due to the
poorer spatial resolution of those early surveys and it is likely they detected
the masing source associated with SFO\,30 and IRAS\,18159$-$1346.

Recently, \citet{healy04} performed several VLA pointings towards M\,16,
concentrating mainly on the dusty pillars. A total of 8 water masers were
detected: a single component towards column II, one component in column IV
likely associated with the driving source of HH\,216 (Section\,\ref{hh216}),
three components in column V and a further 3 components at the location of
SFO\,30 (Section\,\ref{sfo30}). Typical separations between water masers and the
YSOs that excite them are of the order of tens to hundreds of AU for low-mass
protostars \citep[e.g.][]{furuya00} and $\la$\,10$^{4}$\,AU for massive
protostars \citep[e.g.][]{tofani95,foster00}. For the locations with several
maser detections, the physical separation between the components is at least
10$^{4}$\,AU, implying that the masing activity is associated with multiple
protostars. These observations further reinforce the notion that star formation
is ongoing in several locations in the Eagle Nebula.

\section{Individual Objects of Particular Interest}
\label{individual}

\subsection{The Herbig-Haro Object HH\,216}
\label{hh216}

\begin{figure}[t]
\begin{center}
\includegraphics[scale=0.50]{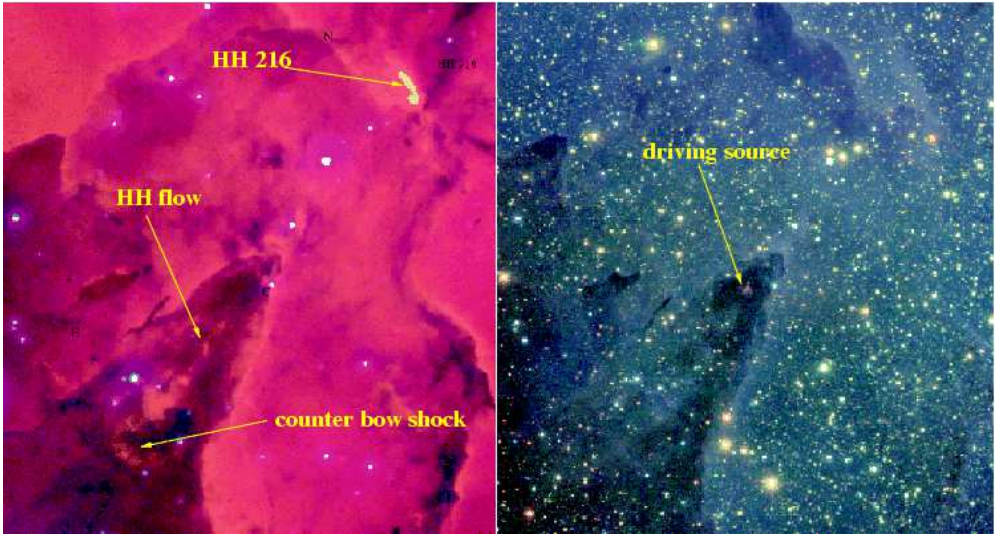}
\end{center}
\caption{Colour composite images of HH\,216 and column\,IV \citep{andersen04}.
Left: H$\alpha$ (red), [S\,{\sc II}] (green) and continuum (blue); Right:
true-colour JHK image. The optical composite shows the positions of HH\,216, the
counter bow shock and the HH flow. The IR composite shows the possible location
of the driving source (see text).}
\label{hh}
\end{figure}

Herbig-Haro (HH) objects are optically bright, shock-excited nebulae powered by
outflows from YSOs (see \citealt{reipurth01} for a review on HH objects) and
thus also act as signposts for embedded star formation occurring nearby. The HH
object in M\,16 (see Figs.\,\ref{trunks_ir} and \ref{hh}) was first identified
by \citet{meaburn82b} near what is now known as column IV; initially labelled
M16-HH1 it was later renamed HH\,216 \citep{reipurth99}. Several authors report
optical and UV spectroscopic observations of this object
\citep{meaburn82a,meaburn84,meaburn90}.

\citet{andersen04} have identified what appears to be the counter bow shock
to HH\,216 (Fig\,\ref{hh}), based on optical line emission, CO and CS emission,
near-IR imaging and gas kinematics. While HH\,216 is redshifted with a radial
velocity of 150\,km\,s$^{-1}$, the new emission-line object is blueshifted with
a radial velocity of $-$150\,km\,s$^{-1}$. A string of optical or IR emission
knots lies between these two objects. A small near-IR extended nebula,
symmetrically placed between the two HH objects, points at the location of the
embedded driving source of the HH system. Water maser emission has also been
detected associated with this object (see previous section).

\citet{linsky07} found a weak X-ray source at the location of HH\,216. This
source has a mean photon energy of the order of 1.9\,keV and an X-ray luminosity
of $\sim$\,10$^{30}$\,erg\,s$^{-1}$, comparable to other HH objects. It is
thought that X-ray emission in these objects originates from mechanical heating
between the jet and the circumstellar medium \citep[e.g.][]{bonito04}. No X-ray
emission was detected towards the driving source of the HH\,216 outflow.

\subsection{Massive Embedded YSOs}
\label{ysos}

The tips of columns I and II (Fig.\,\ref{trunks_ir}) and column V
(Fig.\,\ref{spire_fig}) show evidence of bright YSOs, first identified in IR
images \citep{hillenbrand93}. The most recent analysis was performed by
\citet{indebetouw07} who used Spitzer 3.6$-$24\,$\mu$m photometry to constrain
the Spectral Energy Distribution (SED) of several YSOs. Some of those YSO
sources have been subject to close scrutiny
\citep{thompson02,sugitani02,mccaughrean02,linsky07,sugitani07,indebetouw07} and
are here described in more detail.

\subsubsection{YSO M16\,ES$-$1:}

YSO M16\,ES$-$1 \citep{thompson02} is a very bright and red IR source at the tip
of column I (YSO1 in \citealt{mccaughrean02}; P1 in \citealt{sugitani02}). It
was also detected at mid-IR wavelengths by MSX \citep{price01} and ISO
\citep{pilbratt98}. \citet{thompson02} integrated the object's near-, mid- and
far-IR fluxes to estimate a luminosity $L_{\rm bol}$\,$\sim 200$\,L$_{\odot}$,
indicative of a 4$-$5\,M$_{\odot}$ ZAMS star. Assuming that the observed $J$ and
$H$ magnitudes are photospheric, \citet{mccaughrean02} find that these fluxes
are consistent with a deeply embedded ($A_{\rm V} \sim 27$\,mag),
10\,M$_{\odot}$ ZAMS star. However, \citet{thompson02} find no Pa$\alpha$
emission at the location of this source, suggesting that it cannot be a single
ZAMS object (not even a late B spectral type object). Thus, this object could
either be a small cluster of low-mass PMS stars or an early protostellar source.
Its large IR excess \citep{mccaughrean02} suggests an embedded YSO.
\citet{fukuda02} report the presence of 2.7\,mm continuum emission at the
position of ES$-$1 that they suggest is due to dust emission, even though it
could also be free-free emission from an ultracompact H\,{\sc ii} region.

\citet{sugitani07} identifies strong polarized emission (in the H- and K-bands)
asymmetrically distributed north and south of ES$-$1. They suggest this emission
originates from the walls of two cavity lobes, which were created by the
molecular outflow from the central object. A clear gap in polarization intensity
is seen between the two lobes, that they propose corresponds to a disk-like
structure (previously suggested by \citealt{sugitani02}), perpendicular to the
axis of the cavity lobes. Furthermore, the disk appears to be tilted so that
higher polarization intensity is produced at the north lobe, while higher degree
of polarization is detected at the south lobe.

\citet{linsky07} identify ES$-$1 with a very bright and hard X-ray source. They
estimate $L_{\rm X} \sim 1.6 \times 10^{32}$\,erg\,s$^{-1}$,
$L_{\rm X}/L_{\rm bol} \sim 2.1 \times 10^{-4}$, mean photon energy
\={e}\,=\,3.3\,keV and plasma temperature of $\sim$ 2.2\,keV. These properties
are consistent with those observed for other young, magnetic O-stars (heating by
magnetically channelled wind shocks), but very different from typical O-stars
with X-ray emission produced in weak shocks. Thus they conclude that ES$-$1 is
most likely a magnetically active, high-mass YSO.

\citet{indebetouw07} analysed Spitzer photometry for ES$-$1. The 2$-$24\,$\mu$m
spectral index and mass accretion rate
($1-70\times 10^{-5}$\,M$_{\odot}$\,yr$^{-1}$) are consistent with a Class\,I
source. These authors estimate a bolometric luminosity of
44\,$\pm$\,10\,L$_{\odot}$. Using fits to the object's SED, they estimate a mass
of 4.5\,M$_{\odot}$, with a total extinction (foreground and circumstellar) of
$A_{\rm V} \sim 40$\,mag.

\subsubsection{YSO M16\,ES$-$2:}

YSO M16\,ES$-$2 \citep{thompson02} sits at the tip of column II (YSO2 in
\citealt{mccaughrean02}; T1 in \citealt{sugitani02}). Compared to ES$-$1, this
source is less luminous ($ L_{\rm bol} \sim 20 L_{\odot}$,
\citealt{thompson02,indebetouw07}) and less obscured ($A_{\rm V} \sim 15$\,mag,
\citealt{mccaughrean02}), with an estimated mass of 2$-$5\,M$_{\odot}$
\citep{thompson02,mccaughrean02,indebetouw07}. Based on its near-IR excess
\citep{sugitani02} and mid-IR spectral index \citep{indebetouw07}, ES$-$2 has
been classified as a Class\,II object, i.e., probably at a more evolved stage
than ES$-$1. Water maser emission detected at the tip of column II is not
associated with ES$-$2 \citep{healy04}. Similarly to ES$-$1, \citet{sugitani07}
detect polarized emission at the position of ES$-$2 and propose the presence of
a tilted disk-like structure around this object.

ES$-$2 is the weakest of the X-ray sources considered by \citet{linsky07}. They
estimate $L_{\rm X} \sim 1.26 \times 10^{30}$\,erg\,s$^{-1}$ and
$L_{\rm X}/L_{\rm bol} \sim 1.6 \times 10^{-5}$. They note that the X-ray
luminosity of ES$-$2 is consistent with similar mass young objects in Orion, but
the luminosity ratio is rather low. However, these estimates are very uncertain
due to the small number of detected counts (5 raw counts). Still, it is
possible that this object is too young to be significantly X-ray active,
similarly to the 4 massive EGGs mentioned in Section\,\ref{xrays}.

\subsubsection{Massive YSOs in Column\,V:}
\label{ysos_5}
\citet{indebetouw07} has constructed the SEDs of 2 embedded sources located in
column V. P5\,A is located at the tip of this column and it is marginally
resolved in the Spitzer images, with two IR components corresponding to two of
the three masing sources identified by \citet{healy04}. This source has a
high spectral index, a luminosity of $\sim$\,250\,L$_{\odot}$ and it is very
bright at 24$\mu$m, consistent with it being a young intermediate mass
($\sim$\,6\,M$_{\odot}$) YSO. The object identified as P5\,B sits at the base of
the column; it is less luminous ($\sim$200\,L$_{\odot}$) than P5\,A but it is
also likely to be an embedded YSO.

\subsubsection{IRAS~18152$-$1346:}
\label{myso}

IRAS~18152$-$1346 is the most luminous YSO identified so far in M\,16 and is
located to the west of the main pillars, in a little studied region. This IR
source is also associated with water maser emission \citep{braz83,codella95}.
Fits to the SED suggest that this object has a luminosity of
$\sim$\,1000\,L$_{\odot}$ and a mass of 8\,M$_{\odot}$ \citep{indebetouw07}.

\begin{figure}[h!]
\begin{center}
\includegraphics[scale=0.45]{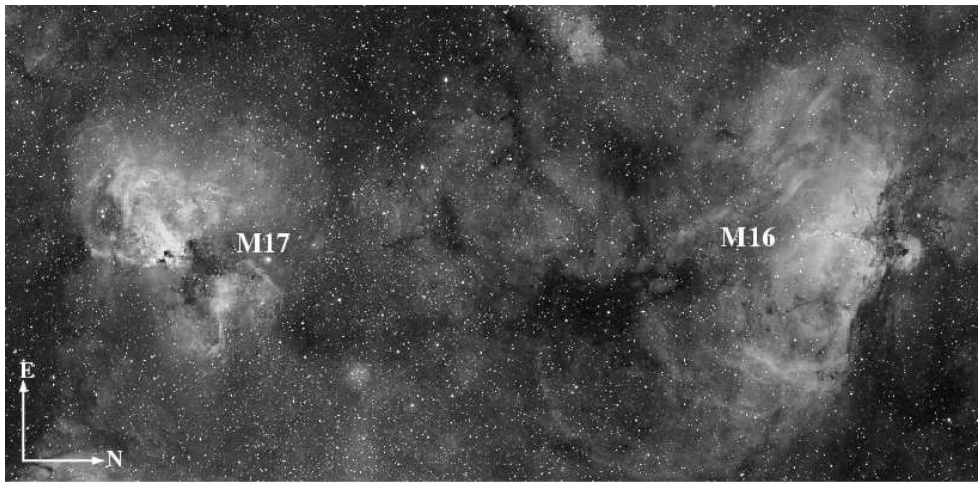}
\end{center}
\caption{M\,16 and M\,17 might be part of the same star-forming cloud complex in
the Sagittarius arm (H$\alpha$ image courtesy of Russell Croman).}
\label{m1617}
\end{figure}

\section{A Link between M\,16 and M\,17}

Both M\,16 and M\,17 (the Omega Nebula) are large star forming regions in the
Sagittarius spiral arm \citep{georgelin70}, at distances from the Sun of
2.1\,kpc for M\,17 (see chapter by Chini \& Hoffmeister) and 1.8\,kpc for M\,16.
Their angular separation is about 2.5$\deg$ (Fig.\,\ref{m1617}).

Based on $^{12}$CO emission maps \citet*{elmegreen79} show that there is
tenuous molecular material connecting M\,16 and M\,17 and region III (a complex
located to the southwest of M\,17). This suggests that the whole region is
a quasi-continuous molecular structure, within which these massive individual
clouds have condensed and begun to form stars \citep[see also][]{moriguchi02}.
\citet*{sofue86} further propose that NGC\,6604 (located to the north of
NGC\,6611) is also part of this ``long string of beads'' of star forming
activity along the Sagittarius arm. They suggest that it also defines an age
sequence: the older region NGC\,6604 (age $\sim$\,4\,Myr, see chapter by
Reipurth) triggers star formation in M\,16 (age $\sim$\,2\,$-$\,3\,Myr), then
M\,17 (age $\sim$\,1\,Myr), via stellar-wind or supernova-blown bubbles.
Alternatively, these regions are a natural progression of star formation as
molecular clouds collapse and form stars due to the passage of the spiral arm.

These regions might actually be part of an even larger star formation complex
\citep{stalbovskii81} that includes also M\,8 (the Lagoon Nebula) and M\,20 (the
Trifid Nebula), extending along the Sagittarius arm at $l = 13\deg \pm 4\deg$.
The location and typical scales of this super-star forming region in the spiral
arms of our Galaxy are consistent with what is observed in other spiral
galaxies, offering us a route to understand the mechanisms for star formation at
large (galactic) scales in the Milky Way.

\vspace{0.5cm}

{\bf Acknowledgements.}  JMO acknowledge financial support from the UK
Science and Technology Facilities Council. I would like to thank all
authors that have contributed with figures to this manuscript.

%%% THE BIBLIOGRAPHY
%%%
%%% CONSULT SECTION 3 OF "INSTRUCTIONS FOR AUTHORS" FOR HOW TO USE NATBIB.
%%% AUTHORS ARE ENCOURAGED TO USE EITHER THE "THEBIBLIOGRAPY" ENVIRONMENT
%%% BY UNCOMMENTING (DELETING THE "%" SYMBOL) THE COMMANDS BELOW, OR BY
%%% USING THE BIBTEX ENVIRONMENT. TO FIND OUT WHICH IS APPLICABLE TO YOUR
%%% CONTRIBUTION, CONSULT THE VOLUME EDITORS FOR YOUR PROCEEDINGS.
%%%

\end{document}